\documentclass[10pt,twocolumn,letterpaper]{article}

\usepackage{cvpr}              

\usepackage[dvipsnames]{xcolor}

\definecolor{cvprblue}{rgb}{0.21,0.49,0.74}
\usepackage[pagebackref,breaklinks,colorlinks,citecolor=cvprblue]{hyperref}
\usepackage{dblfloatfix}
\def\paperID{5626} 
\def\confName{CVPR}
\def\confYear{2024}

\title{EXACT-Net: EHR-guided lung tumor auto-segmentation for non-small cell lung cancer radiotherapy}

\author{Hamed Hooshangnejad\\
Johns Hokins University\\
Baltimore, MD\\
{\tt\small hamed@jhu.edu}
\and
Xue Feng\\
Carina Medical\\
Lexington, KY\\
{\tt\small xfeng@crainaai.com}
\and
Gaofeng Huang\\
Carina Medical \\
Lexington, KY\\
{\tt\small ghuang@crainaai.com}
\and
Rui Zhang\\
University of Minnesota \\
Minneapolis, MN\\
{\tt\small zhan1386@umn.edu}
\and
Katelyn Kelly\\
Johns Hokins University\\
Baltimore, MD\\
{\tt\small kkelly80@jhmi.edu}
\and
Quan Chen\\
Mayo Clinic \\
 Phoenix, AZ\\
{\tt\small quanchen@gmail.com}
\and
Kai Ding\\
Johns Hopkins University \\
Baltimore, MD\\
{\tt\small kding1@jhmi.edu}
}

\begin{document}
\maketitle
\nolinenumbers
\begin{abstract}
Lung cancer is a devastating disease with the highest mortality rate among cancer types. Over 60\% of non-small cell lung cancer (NSCLC) patients, which accounts for 87\% of diagnoses, require radiation therapy. Rapid treatment initiation significantly increases the patient's survival rate and reduces the mortality rate. Accurate tumor segmentation is a critical step in the diagnosis and treatment of NSCLC. Manual segmentation is time and labor-consuming and causes delays in treatment initiation. Although many lung nodule detection methods, including deep learning-based models, have been proposed, there is still a long-standing problem of high false positives (FPs) with most of these methods. Here, we developed an electronic health record (EHR) guided lung tumor auto-segmentation called EXACT-Net (EHR-enhanced eXACtitude in Tumor segmentation), where the extracted information from EHRs using a pre-trained large language model (LLM), was used to remove the FPs and keep the TP nodules only. The auto-segmentation model was trained on NSCLC patients' computed tomography (CT), and the pre-trained LLM was used with the zero-shot learning approach. Our approach resulted in a 250\% boost in successful nodule detection using the data from ten NSCLC patients treated in our institution.
\end{abstract}    
\section{Introduction}
\label{sec:intro}

Lung cancer is the deadliest cancer type, with more than 238,000 new diagnoses each year in the US \cite{Siegel2023}. Radiation therapy (RT) is the common and preferred treatment modality for medically inoperable non-small cell lung cancer (NSCLC) \cite{Cao2019}, accounting for 87\% of lung cancer cases \cite{Tandberg2018}. Over 60\% of NSCLC diagnoses, more than 142,000 patients, require RT at least once over the course of their disease \cite{Herbst2018,Tyldesley2001}. However current RT workflow is time-consuming and comprises numerous steps, resulting in a considerably long tie to treatment initiation (TTI).

Previous studies have reported that lung cancer mortality rate increases with prolonged TTI \cite{Hanna2020, Samson2015, Khorana2019,Cushman2021,Hooshangnejad2023, Hooshangnejad2023a} , not only increasing the chance of new lymph node involvement, site of disease, and but also re-staging\cite{Mohammed2011}, demonstrating that reducing the TTI is critical for patient survival and treatment outcome.

\begin{figure*}
	\centering
	\includegraphics[width=1\linewidth]{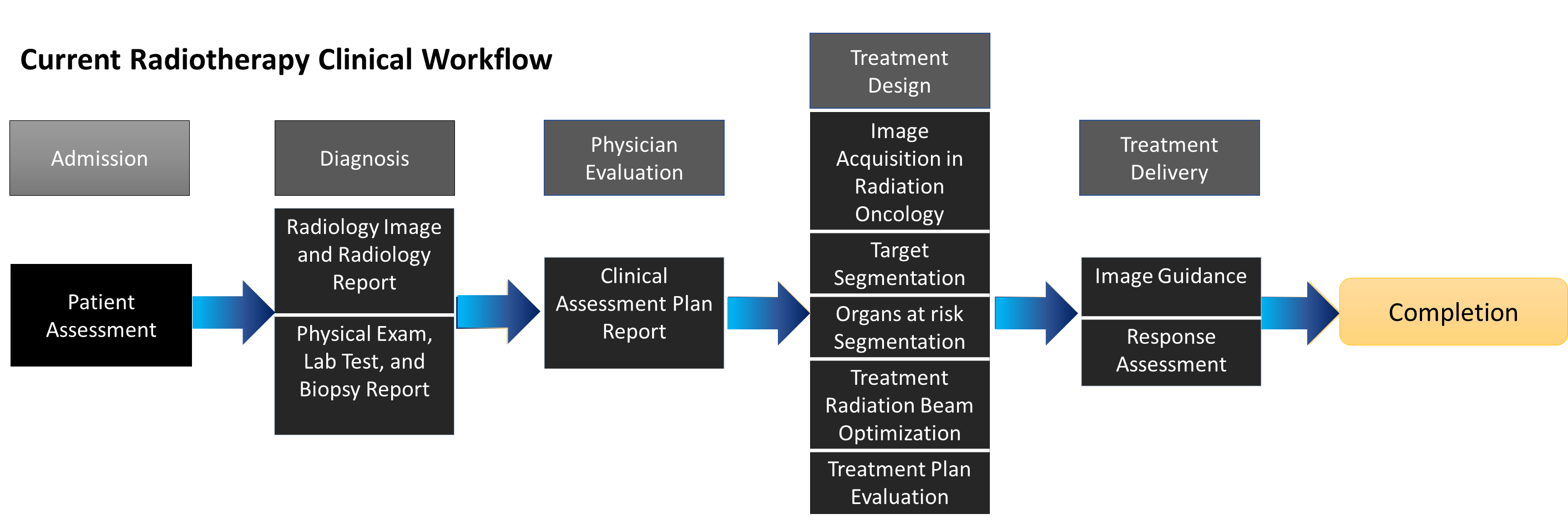}
	\caption{The current radiotherapy workflow, consisting of multiple steps including the target and organs at risk segmenation.}
	\label{fig:f1}
\end{figure*}

Tumor segmentation is a vital step in the diagnosis and treatment workflow. Figure \ref{fig:f1} shows the current radiotherapy workflow, which starts with patient registration and assessment. The patient is then sent for radiology diagnostic image acquisition and pathology tests. The radiologist and pathologist provide their initial diagnosis based on the test as a report for the physician’s evaluation. Based on the physician's clinical evaluation and radiology and pathology reports, a clinical assessment plan report is generated that not only includes the diagnostic reports but also the concluding diagnosis. To design the treatment plan a new patient's CT scan is acquired and then the pivotal step of target and organs at risk segmentation is carried out, which is critical for radiotherapy plan optimization and effectiveness of treatment. Treatment is delivered with image guidance and the treatment response is assessed for potential amendment. If no further treatment is required, the treatment will be concluded.

The main imaging modality used for diagnosis and RT treatment is Computed Tomography (CT). Since these CT scans are very high quality with thin slices, they are large volumetric scans with millions of voxels, making the diagnosis and tumor/target segmentation challenging and time-consuming even for radiation oncologists \cite{Le2023}. Given that manual segmentation is labor-intensive and considerably time-consuming, an automated lung cancer nodule segmentation method is extremely desirable. 

Although there have been numerous efforts for automatic lung nodule segmentation, a major and long-lasting issue with the automatic methods is their high false positive (FP) rate. As a result, most algorithms are comprised of nodule detection and FP reduction system\cite{Le2023,Gruetzemacher2018,Khosravan2018, Zhao2022}, or are dealt with by manual cropping prior to nodule segmentation \cite{Zhao2022}. Pulmonary blood vessels, lung borders, and noise from CT scanners may result in high false-positive rates of detected nodules \cite{Yu2023}. Different methods such as classic machine learning, feature extraction \cite{Shamas2023}, and deep learning (DL)\cite{Li2022} algorithms have been used for FP reduction but there are still challenges with these methods.

\begin{figure*}[!b]
	\centering
	\includegraphics[width=1\linewidth]{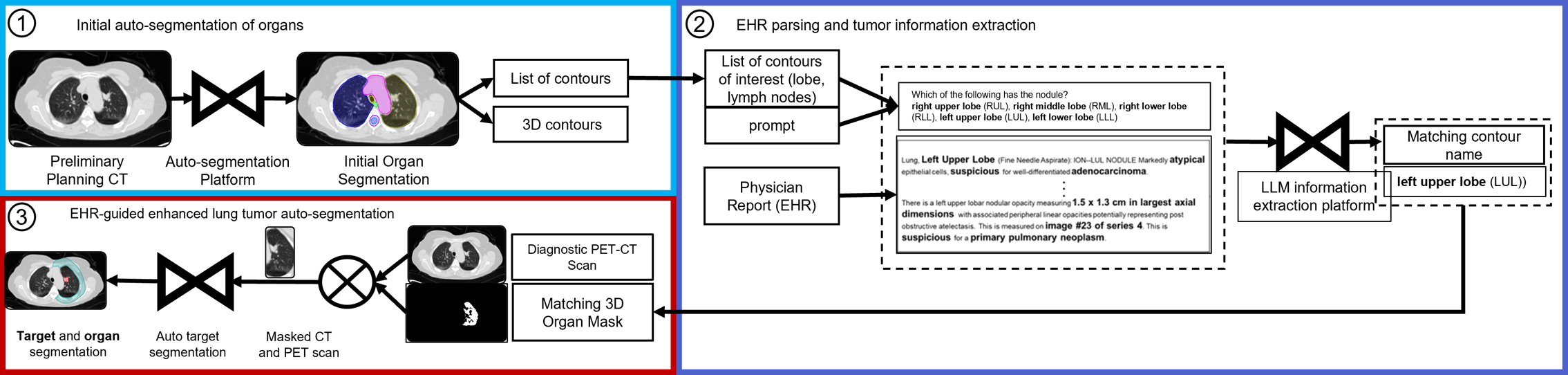}
	\caption{Overview of EHR-guided automated target segmentation system. The auto-contouring platform contour the initial structures on the diagnostic CT scan. Due to advantages of PET scan for improved target segmentation, it will be used as the second primary imaging modality for target segmentation platform. The NLP based algorithm will extract critical information regarding the location and shape of tumor.}
	\label{fig:f2}
\end{figure*}

Meanwhile, information pinpointing the location of the lung tumor is already available in the patient’s electronic health record (EHR) such as clinical decisions outside the traditional simulation CT (such as reports from diagnostic radiology scans and pathology reports). Still, little attention has been paid to parsing and incorporating this data into target segmentation. These reports have been meticulously written and contain valuable information regarding the location, shape, and size of the tumor.

Recently, Large language models (LLMs) have shown remarkable performance in a variety of natural language processing (NLP) tasks, such as summarization, question, and answering\cite{Fan2023}. Using neural networks with several billion parameters, and self-surprised learning on a large corpus of unlabeled text data, LLMs can very efficiently parse information and generate human-like responses to prompts. OpenAI GPT3 and 4, Google Bard\cite{Siad2023}, and Meta LLAMA are only a few examples of state-of-the-art LLMs.

ChatGPT is developed from InstructGPT\cite{Ouyang2022}, a fine-tuned version of GPT3, by fine-tuning for dialog interface. While InstructGPT is designed for processing prompts to provide responses to pre-defined tasks and instructions, ChatGPT is designed to engage in conversion and create more naturalistic responses\cite{Ouyang2022}.  ChatGPT is trained on more than 300 billion words and has more than 100 trillion parameters, showing remarkable capability in generating and collaborating with users for creative and technical downstream tasks. Recent findings have shown that with in-context zero-shot and few-shot leanings, LLMs can adapt to many novel downstream tasks \cite{Agrawal2022,Kojima2022}.

“Prompt Engineering” is an important aspect of zero-shot and few-shot learning, which is best defined as providing effective contextual cues to help the LLMs with the given task. More recently, LLMs have also been used to produce clinical reports\cite{Shachar2022,Biswas2023} or, more focused, generate domain-specific text, such as radiation oncology treatment regimens\cite{Liu2023a}. 

In this work, we demonstrate the feasibility of an EHR-guided lung tumor auto-segmentation method (Figure \ref{fig:f2}). To deal with the critical issue of FPs in lung nodule segmentation, LLM-parsed patient-specific EHR is used to aid the tumor auto-segmentation method eliminate the FP nodules from the result. This is because the main tumor to be treated can be one of the detected nodules and not all lung nodules are the cancerous tumor. Thus, without relying on direct user input, FPs were significantly reduced, and automatic target segmentation was improved. Specifically, the LLM-extracted tumor location served as an input bounding area for the CT images, to guide the segmentation network.

\section{Related Work}

Lung cancer is a devastating disease, and tumor segmentation is a critical part of radiotherapy treatment practice, as initiating the treatment depends on the tumor and organs at risk segmentation. Thus, the lung tumor segmentation has been of much interest to many researchers over the years. The first attempts for lung tumor segmentation were based on image characteristics such as shape, intensity, and texture. These methods were computationally expensive and hard to generalize to difficult cases and were mainly based on traditional image processing methods such as intensity and adaptive thresholding, image registration, and region growing \cite{Kenneth1996,Uzelaltinbulat2017}.

CT scan, amongst all the other imaging modalities, is the standard of care for cancer therapy such as radiation therapy\cite{Delbeke2006}, the diagnostic CT scan, which is acquired for tumor delineation, has the field of view that is bound to the lungs to make lung parenchyma easily distinguishable. However, even with this field of view tumor only accounts for a small number of voxels compared to the entire CT, making lung tumor segmentation very challenging \cite{Mercieca2021,VanMeerbeeck2011}.

Convolutional neural networks (CNNs) have been used in numerous medical image processing fields, such as segmentation, classification, detection, and reconstruction, and have achieved great success. CNNs have been applied to lung CT scans for lung pathology segmentation \cite{Zotin2019}, lung volume segmentation \cite{Skourt2018}, and lung region classification \cite{Zotin2019, Asuntha2020}.  In the case of entire lung volume segmentation, there are two common approaches: (i) 2D segmentation, in which each CT slice is segmented independently. In this case, the CNN architectures such as UNet or VGG-16, are used to segment each slice of CT in a 2D manner\cite{Asuntha2020,Patil2020,Pawar2021}. Generative adversarial networks (GAN) have also been used for segmentation, where the GAN model is used to encode features of CT slice and then the encoder-decoder is used to segment lung volume\cite{Alom2019}; (ii) 3D segmentation, in which the CT scan is treated as 3D volume, and is segmented for lung in 3D. Previously, V-net and more improved networks like RU-Net, and R2U-Net have been used for volumetric segmentation\cite{Milletari2016,Negahdar2018,Xie2019}.

In the case of lung tumor segmentation, the common practice is a 2-step framework in which a delineation model first segments all suspicious tumors and an FP reduction model reduces the FPs. For instance, Xie H. et al.\cite{Xie2019} used a 2D Faster R-CNN model with deconvolutional layers to magnify the feature maps to detect all the candidate nodules from slices of CT scan. Then, they trained a classifier to reduce FPs.  Another work used a different architecture, ResNet\cite{Zhang2020}, to segment tumors for NSCLC cases. For 3D tumor segmentation, Kopelowits E., et al. \cite{Kopelowitz2019} used MaskRCNN to detect 3D nodules on CT scans, and Kamal U., et al.\cite{Kamal2020} used a Recurrent 3D-DenseUNet for lung tumor segmentation. More recently, Le V. and Saut O. \cite{Le2023} used a variant of the UNet model, RRc-UNet 3D, which is augmented with residual recurrent block, for 3D segmentation of NSCLC tumors.

\section{Methods and Materials}
Here we have developed a 2-step framework for an EHR-guided tumor auto-segmentation, where an LLM model extracts tumor location, tumor size, and lymph node involvement from clinical reports  (Figure \ref{fig:f2}) , and the parsed data from LLM such as tumor location and size will be used to refine the result and a robust target delineation. 

\subsection{LLM with Prompts for Tumor Phenotype Extraction}
To demonstrate the feasibility of EHR-guided tumor segmentation, we used the GPT 3.5 Turbo LLM through OpenAI API. We used an in-house Python interface to the API to read the EHR text files and feed them to the model. To access the API, an API key is required that can be obtained from OpenAI official website. 

An important step is setting up the appropriate parameters for the LLM before submitting the prompts. Especially, “Temperature” is an important parameter determining the certainty vs creativity of the model. In our case, maximum certainty is desirable, to which we used zero for the temperature.  

To demonstrate the feasibility, the pathology lab and radiology image reports and CT images of 10 cases of lung cancer patients treated with RT at Johns Hopkins Hospital were used for this study. As, at the moment, we did not have the resources to deploy the local model, we manually de-identified all the patient data, removed the names and locations, and altered the dates of the procedures and visits.

We implemented our novel strategy using the zero-shot learning approach meaning that we have not fine-tuned the model for cancer reports. Thus, an important step is “prompt engineering”, which is the careful design of the prompts that work with ChatGPT to help and guide the pre-trained GPT3.5 model in extracting the relevant information. A prompt is an instruction set that regulates the LLM’s capabilities and affects the generated outputs. Prompts not only can filter information but also create new interaction paradigms, such as directing the LLM to search for relevant information that has not been seen before. 

Here, we also used prompts as a tool for LLM self-adaptation to new tasks. For instance, we can extract the  tumor location by using the prompts such as "\textit{find the current lung lobe that the determinate tumor/carcinoma/malignancy is involving in this report:}" or "\textit{based on the report, where is the tumor or carcinoma located?} followed by possible options of \textit{right upper lobe (RUL), right middle lobe (RML), right lower lobe (RLL), left upper lobe (LUL), left lower lobe (LLL)}.  Currently, in the medical domain  automatically learned prompts are less common as they are often not human-readable\cite{Liu2023}. As interpretably is essential, we manually engineered prompts with trial and error.

\begin{figure*}
	\centering
	\includegraphics[width=0.7\linewidth]{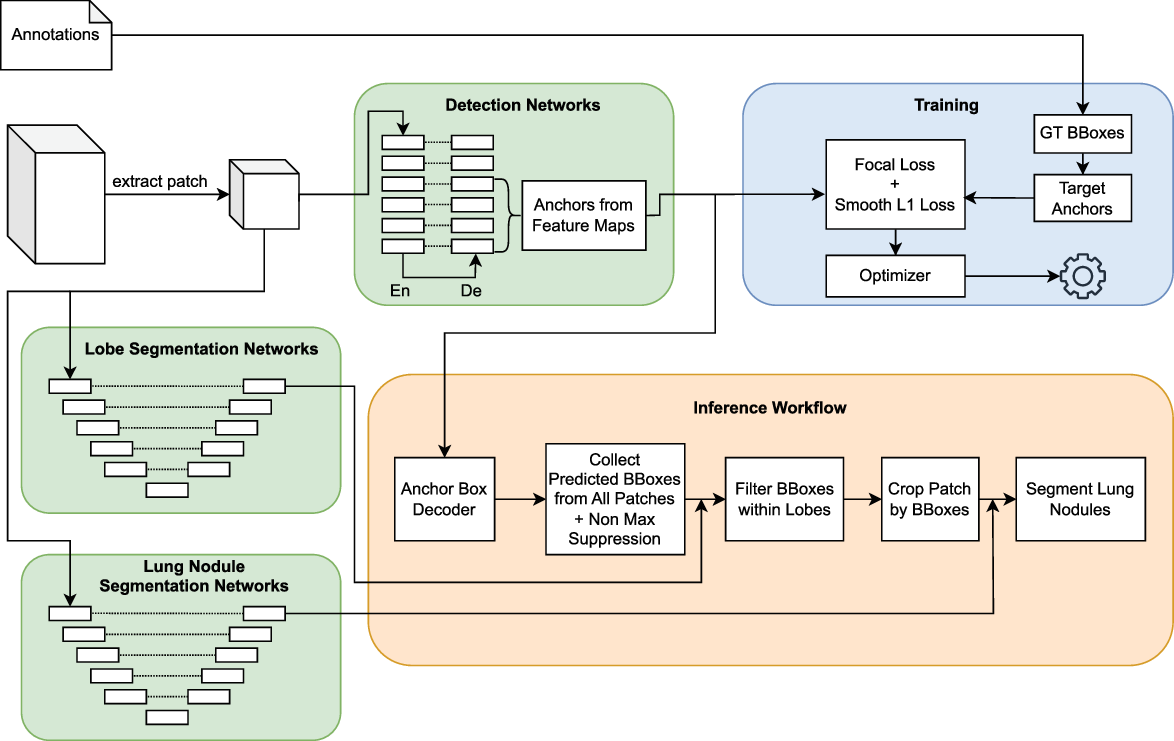}
	\caption{Detailed tumor auto-segmentation model architecture}
	\label{fig:f3}
\end{figure*}

\subsection{Tumor Auto-segmentation Algorithm Design and Training}
To automatically detect and segment the lung nodules, we developed an automated deep-learning model.

\subsubsection{Architecture}
We chose to employ the UNet3D architecture for our 3D segmentation networks (Figure \ref{fig:f3}). This selection was based on the architecture's well-established effectiveness in segmenting 3D medical images\cite{Feng2020,Lin2017}. 2D object detection in computer vision has significantly advanced and matured. When it comes to 3D object detection in medical images, recent efforts have drawn inspiration from the progress in 2D single-stage detectors, particularly RetinaNet\cite{Lin2017}, and leveraged the successful 3D segmentation models, UNet3D, to construct a 3D detection framework termed as Retina-UNet3D.

RetinaNet stands out due to its feature pyramid network (FPN), which efficiently handles object detection across various scales. This FPN-based approach serves as the fundamental building block for our 3D object detection framework, providing a strong foundation for our work in the medical image domain. In the context of 3D segmentation, the UNet3D encoder-decoder approach has proven highly effective. Leveraging this success, Retina-UNet3D adopts the UNet3D encoders and establishes connections between the encoder outputs and the detection decoder heads as the UNet architecture also shares a similar pyramid network design.

This design choice aligns with the principles of both FPN and UNet, allowing us to leverage the benefits of feature pyramid structures for our 3D detection framework while building upon the successful foundation of UNet3D in segmentation tasks.

\subsubsection{Loss Functions}

In our segmentation tasks, we used a dual loss strategy: 

\begin{enumerate}
\item Categorical Cross-Entropy Loss: This loss is employed for pixel-level classification, ensuring precise object boundary delineation and class-specific region identification.
\begin{equation}
	L\left(p,q\right)=-\sum_i{y_i}{\mathrm{log} \left(p_i\right)\ }
\end{equation}

\item Dice Loss: Dice loss is used for image-level Intersection over Union (IoU) measurement, assessing overall segmentation quality.

\begin{equation}
L\left(y,p\right)=1-\frac{2\sum_i{y_ip_i}}{\sum_i{y^2_i}+\sum_i{p^2_i}}
\end{equation}

\end{enumerate}

These two losses are in same order of magnitude, so we simply added them as the final dual loss. This approach stroked a balance between pixel-wise accuracy and image-level fidelity, enhancing segmentation performance across both detailed object recognition and holistic quality assessment.

In our detection tasks, we used smooth L1 loss\cite{ArmatoIII2011} for box regression heads and focal loss \cite{Girshick2015} for box classification heads. Hereby, we used $\delta$= 1. MAE is the mean absolute error between predicted bounding boxes and ground-truth bounding boxes.

\begin{equation}
	smoothL1 =\left\{ \begin{array}{cc}
		\frac{0.5\ {\left(MAE\right)}^2}{\delta }\ \ \ \ \ if\ \ MAE<\delta  \\ 
		MAE-0.5\  \delta \ \ \ \ \ \ \ otherwise \end{array}
	\right.
\end{equation}

\begin{equation}
	FL\left(p,y\right)=\left\{ \begin{array}{c}
		-\alpha {\left(1-p\right)}^{\gamma }{\mathrm{log} \left(p\right)\ }\ \ \ \ \ \ \ \ if\ y=1 \\ 
		-\left(1-\alpha \right)p^{\gamma }{\mathrm{log} \left(1-p\right)\ if\ y=0\ \ } \end{array}
	\right.
\end{equation}

\subsubsection{Dataset and Data Preprocessing}
We used the publicly available Lung Image Database Consortium imaging collection (LIDC-IDRI) from TCIA \cite{Tang2019} as it contains high-quality annotations for all nodules $>= 3mm$ with consensus from experienced radiologists. With the same LIDC-IDRI dataset, \cite{Feng2019} annotated 50 CT scans on lobe segmentation.

We standardized the voxel spacing of all CT datasets to 1mm × 1mm × 1mm. In addition, we applied intensity clipping, constraining the Hounsfield Units (HU) within the range of -1000 HU to 600 HU. This adjustment was made to prevent the model from overly emphasizing irrelevant information.

Furthermore, in preparation for training our detection model, data containing segmentation contours were transformed into bounding boxes. This conversion facilitated the training process and enabled effective detection.

\subsubsection{Data Augmentation}
For our segmentation task, we employed a comprehensive data augmentation approach, which included 3D random rotations within a range of -30 to 30 degrees along each axis, random scaling between 70\% and 140\%, random cropping, random flipping along each axis, random Gaussian noise, random Gaussian blur, and random variations in brightness, contrast, and gamma. 

For our detection task, we used the same augmentation methods as segmentation but removing rotation and flipping approaches. This combined augmentation strategy was instrumental in enhancing the robustness and generalization of our models.

\begin{figure*}
	\centering
	\includegraphics[width=1\linewidth]{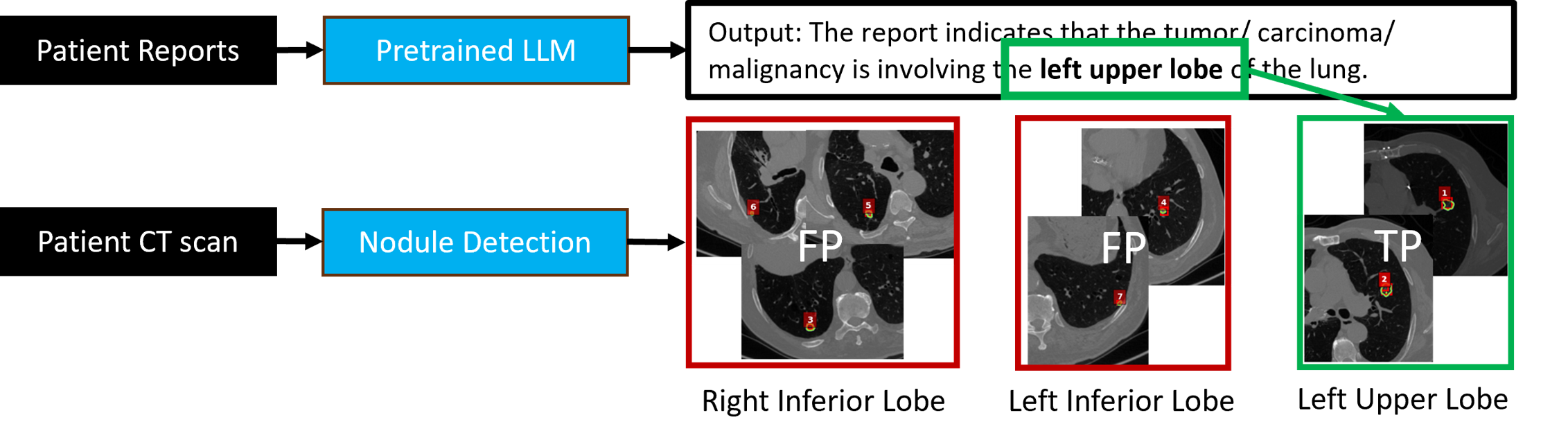}
	\caption{Showing an example case of EHR-guided tumor auto-segmentation where the extracted information regarding the confirmed tumor is used to remove FPs.  The nodule detection algorithm detected and classified seven nodules as malignant for this sample case (Case ID 4). Out of the seven nodules, three were found in right inferior lobe (RIL) and two were found in left inferior lobe (LIL) and two in left upper lobe (LUL). According to the EHR extracted information, however, the confirmed tumors were in LUL. Using this information, the result is automatically refined, and the FPs are removed.}
	\label{fig:f4}
\end{figure*}

\subsubsection{Training and Evaluation Methods}
Our physical device was equipped with a Nvidia RTX 4080 with 16GB of memory. To address the considerable memory demands associated with 3D medical images, we adopted a sliding patch method, enabling us to effectively train models on large-sized images\cite{Isensee,Feng2019} .

For the lung nodule detection model, training was conducted over 500,000 iterations, employing patch sizes of 96 × 96 × 96 (equivalent to 96mm × 96mm × 96mm) with a batch size of 1. In the case of the lung nodule segmentation model, it underwent training for 300,000 iterations, with patch sizes set at 64×64×64 (64mm × 64mm × 64mm) and a batch size of 1. 

Similarly, the lobe segmentation model was trained for 300,000 iterations, with patch sizes configured as 96 × 192 × 192 (equivalent to 288mm × 192mm × 192mm, with 3-times down-sampling along the axial axis) and a batch size of 1. These training settings were tailored to accommodate the hardware limitations while ensuring efficient and effective model training.

In the first stage, we employed the lung nodule detection model to generate a list of bounding box candidates for potential nodules. Subsequently, we applied the lobe segmentation model to obtain lobe masks and assigned lobe location information to each bounding box. Any candidate bounding box lacking contact with a lobe was discarded from further consideration.

In the second stage, we cropped 64 × 64 × 64 patches for each remaining candidate. These patches were then processed using the lung nodule segmentation model to delineate the contours of the nodules. This two-stage approach allows us to first identify potential nodules and then precisely segment their contours.

We assessed the quality of segmentation results using the Dice Coefficient Score (DSC), which provides a measure of the overlap between predicted and ground-truth segmentation. 

\begin{equation}
	DSC=\frac{|x\cap y|}{\left|x\right|+|y|}
\end{equation}

For our detection results, we employed the mean Average Precision (mAP) metric with two Intersection over Union (IoU) thresholds at 0.5 and 0.7 for 3D objects. An IoU of 0.5 signifies a good match for 3D small objects and an IoU of 0.7 represents an excellent match. These IoU thresholds allow us to precisely assess the performance of our detection model across varying levels of accuracy and provide a comprehensive evaluation of its object detection capabilities. 

\begin{equation}
	mAP=\frac{1}{N}\sum^N_{c=1}{\frac{TP\left(c\right)}{TP\left(c\right)+FP(c)}}
\end{equation}

As we only had one class of detection object, mAP is equivalent to AP.

\subsection{Experiment Design}
We verified our approach and demonstrated the effect of EHR-extracted information on the tumor segmentation model performance by conducting a comparison study. We used data from 10 test patients from the Johns Hopkins Hospital database, unseen by both the LLM and tumor auto-segmentation model. Once, the nodules were detected and classified using the entire chest CT scan without any EHR guidance. First, the entire body was automatically segmented using the anonymized patients’ diagnostic CT scan for all organs including lung lobes and lymph nodes. Then, auto lung nodule detection and classification software detected and classified all supposedly malignant nodules.  

Next, using the EHR extracted information, we only presented the auto-segmentation model with the confirmed location of tumor according to the clinical reports. For instance, if reports mention that tumor is located in “left upper lobe (LUL)”, the LUL mask was used to mask all other structures and only the LUL was presented to the segmentation model for nodule detection.

\section{Results}
\subsection{Tumor auto-segmentation performance}

\begin{table} 
	\centering
	\caption{Dice score of auto-segmentation method}
	\begin{tabular}{|c|c|}
		\hline 
		Lung Region & DSC \\
		\hline
		RUL & 0.94 \\
		\hline
		RML & 0.83 \\
		\hline
		RLL & 0.96 \\
		\hline
		LUL & 0.98 \\
		\hline
		LLL & 0.97 \\
		\hline
		Lungs Overall & 0.97 \\
		\hline
		Lung Nodule & 0.67 \\
		\hline
	\end{tabular}
	\label{T:1}
\end{table}

The evaluation of segmentation results using the Dice Coefficient Score (DSC) revealed that lobe segmentation closely approximated the manual annotations made by radiologists. This high level of performance in lobe segmentation significantly reinforced the viability of our experiments, which focused on tumor identification based on lobe-specific information obtained from radiology reports.

In comparison to lobes and lungs, lung nodules are notably smaller in size. It's important to note that smaller regions tend to yield lower DSC scores (Table \ref{T:1}), given the inherent challenges in segmenting small structures accurately. Therefore, achieving a DSC score of 0.67 for lung nodule segmentation can be considered a favorable outcome (Table \ref{T:2}).

\begin{table}
	\centering
	\caption{lung nodule detection performance}
\begin{tabular}{|c|c|c|}
	\hline
	& AP@IoU=0.5 & AP@IoU=0.7 \\
	\hline
	Lung Nodule & 65.63 & 59.15 \\
	\hline
\end{tabular}
\label{T:2}
\end{table}

The results were measured in the validation subset of LIDC-IDRI with 201 CTs and 399 solid nodules. Our lung nodule detection model has demonstrated strong performance in identifying solid nodules. However, it's important to note that, in this process, some false positives may arise, and there may be challenges in detecting ground glass nodules.

\subsection{LLM prompt design}

To do so, first, as an “assistant prompt”, we gave a list of lung lobes as options to direct pre-trained LLM to find the lobe with the tumor. Secondly, we used the following as the “user prompt”: "\textit{find the current lung lobe that the determinate tumor/carcinoma/malignancy is involving in this report:}". 

The are three keywords in this prompt to effectively guide pretrained LLM, (i) current, as the patient may have a history of several lung tumors treatment, it is important to specify the currently under treatment tumor; (ii) determinate, the patient reports may include some indeterminate nodules which we are not interested in for current treatment; (iii) tumor/carcinoma/malignancy, we used the common words used in report to refer to tumors in reports so that we do not miss any information regarding tumor location.

Furthermore, to find the malignant lymph nodes, we specifically used the pathology report with the following “user prompt”: "\textit{find out what lymph station/node are malignant in this report:}".

With the above-mentioned prompt, we could find the right tumor location with 100\% accuracy, and it is important to consider that this is based on zero-shot learning which means no fine-tuning has been done on the model. Again, we could find the involved lymph nodes with 100\% accuracy from ten reports.

\subsection{EHR-guided Tumor Auto-segmentation Experiment}

Table \ref{T:3} summarizes the results for the ten test patients. As seen, when the tumor segmentation algorithm is used without any EHR information, in only 20\% of patients the result matches the ground truth, and similar to previous models, there is a false positive issue with initial nodule detection. On the other hand, when the EHR extracted information is used to guide the tumor detection, for 70\% of the patients the patient the result matches the ground truth. Thus, in other words, using medical report information resulted in a huge increase in the number of successful tumor auto-segmentation cases 250\% boost in performance, demonstrating the power of our approach in reducing the FPs.

\begin{table}
	
	\centering
	\caption{The EHR-guided tumor segmentation experiment result}
\begin{tabular}{|p{0.5in}|p{0.5in}|p{0.5in}|p{0.5in}|p{0.5in}|p{0.5in}|}
	\hline
	Case ID & Ground Truth & Detected Nodules & Removed Nodules & Matching Ground Truth \\
	\hline
	1 & 1 & 2 (FP) & 1 & Yes \\
	\hline
	2 & 1 & 1 & 0 & Yes \\
	\hline
	3 & 1 & 1 & 0 & Yes \\
	\hline
	4 & 2 & 7 (FP) & 5 & Yes \\
	\hline
	5 & 2 & 4 (FP) & 2 & Yes \\
	\hline
	6 & 1 & 0 (FN) & 0 & No \\
	\hline
	7 & 2 & 5 (FP) & 3 & Yes \\
	\hline
	8 & 1 & 4 (FP) & 2 & No (FP) \\
	\hline
	9 & 1 & 3 (FP) & 2 & Yes \\
	\hline
	10 & 1 & 1 (FP) & 1 & No \\
	\hline
\end{tabular}
\label{T:3}
\end{table}

Figure \ref{fig:f4} shows a sample case (Case ID 4), where the nodule detection algorithm detected and classified seven nodules as malignant. Out of the seven nodules, three were found in the right inferior lobe (RIL) and two were found in left inferior lobe (LIL) and two in left upper lobe (LUL). According to the EHR extracted information, however, the confirmed tumors were located in LUL. Using this information, the result is automatically refined, and the FPs are removed.

\begin{figure}
	\centering
	\includegraphics[width=1\linewidth]{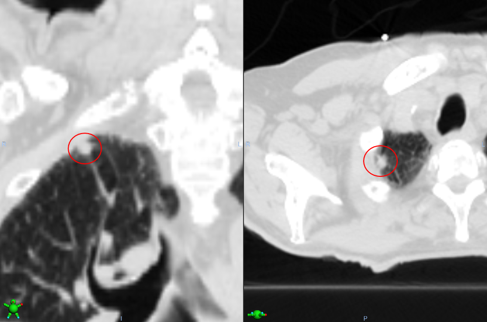}
	\caption{Case 6 where the nodule is very close to chest wall making the nodule detection very challenging. Left showing the coronal view, and right axial view, with tumor marked with red circle.}
	\label{fig:f5}
\end{figure}

\begin{figure}
	\centering
	\includegraphics[width=1\linewidth]{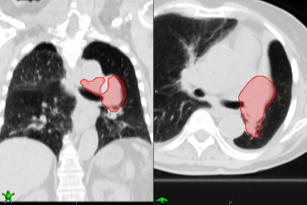}
	\caption{Case 10, our approach averted a false positive in right upper lobe, however, due to highly advanced stage 4 cancer, the nodule detection failed to detect the advanced tumor shown in red shade (left coronal view, right axial view)}
	\label{fig:f6}
\end{figure}

\section{Discussion}

In this study, we presented a novel EHR-guided tumor auto-segmentation method. We showed that incorporating the EHR information into segmentation, as a prior knowledge of confirmed tumor location, is very effective in reducing the FPs. We evaluated the method using the data from ten test cases of actual patients from our institution database.

We are aware that our study may have a few limitations. First, as seen in Table \ref{T:3}, we see two cases (6 and 10) with false negatives (FN), which is highly undesirable. Upon further analysis, we realized that for case 6 the reason that the segmentation model did not detect the nodule is that, (i) as seen in Figure \ref{fig:f5}, the nodule is very close to the lung wall making the nodule detection very challenging, which has also been reported by previous works\cite{Le2023}; (ii) currently, our model is designed as a standalone nodule detector, thus it is restrictive to detect fewer FPs, which resulted in some true positives (TP) to be filtered out. With the EHR extracted information and lowering the classifier threshold, the issue is resolved.

As for case 10 (Figure \ref{fig:f6}), the nodule detection method found an FP nodule in the right upper lobe. According to the EHR report the tumor was located in the left upper lobe, thus the FP was averted. However, case 10, particularly, is highly advanced stage 4 and thus very challenging. As seen in Figure \ref{fig:f6}, the tumor spread significantly and involved a large area of lymph nodes. Because currently our model is not trained on these challenging cases, it failed to detect the tumor. 

\section{Conclusion}

In this study, we presented a novel method for enhancing the nodule detection performance by augmenting the model with EHR-extracted information. EXACT-Net uses a pre-trained LLM model to extract the clinically confirmed tumor location and use that to filter our FPs. We used a zero-shot learning approach by taking advantage of prompt engineering to extract the relevant information. Thus, with EXACT-Net approach we could avert FPs which is the long-standing problem of most nodule detection algorithms.

\section{Acknowledgments}

Research reported in this publication was supported by the National Institutes of Health (award numbers XXX). The content is solely the responsibility of the authors and does not necessarily represent the official views of the National Institutes of Health.

{
    \small
    \bibliographystyle{ieeenat_fullname}
    \bibliography{main}
}


\end{document}